\documentclass[prd,11pt,preprint,nofootinbib,showkeys]{revtex4-1}
\usepackage[mathcal]{euscript}
\usepackage[latin1]{inputenc}
\usepackage{latexsym}
\usepackage{amsmath}
\usepackage{hyperref}
\usepackage{amsmath}    
\usepackage{graphicx}   
\usepackage{verbatim}   
\usepackage{color}      
\usepackage{subfigure}  
\usepackage{hyperref}   
\usepackage{bm}
\usepackage{graphicx}
\usepackage{amssymb}
\usepackage{bbm}
\usepackage{float}
\usepackage{slashed}
\usepackage{amsfonts}
\usepackage{euscript}
\usepackage{amsmath}    
\usepackage{mathrsfs} 
\usepackage{bbding}
\usepackage{pifont}
\newcommand{\cm}[1]{}
\usepackage{cancel}

\makeindex
\begin{document}

\title{Weyl relativity: A novel approach to Weyl's ideas}

\author{Carlos Barcel\'o}
\email{carlos@iaa.es}
\affiliation{Instituto de Astrof\'{\i}sica de Andaluc\'{\i}a (IAA-CSIC), Glorieta de la Astronom\'{\i}a, 18008 Granada, Spain}
\author{Ra\'ul Carballo-Rubio}
\email{raul.carballo-rubio@uct.ac.za}
\affiliation{Department of Mathematics \& Applied Mathematics, University of Cape Town, Private Bag, Rondebosch 7701, South Africa}
\author{Luis J. Garay}
\email{luisj.garay@ucm.es}
\affiliation{Departamento de F\'{\i}sica Te\'orica II, Universidad Complutense de Madrid, 28040 Madrid, Spain}
\affiliation{Instituto de Estructura de la Materia (IEM-CSIC), Serrano 121, 28006 Madrid, Spain}

\begin{abstract}{In this paper we revisit the motivation and construction of a unified theory of gravity and electromagnetism, following Weyl's insights regarding the appealing potential connection between the gauge invariance of electromagnetism and the conformal invariance of the gravitational field. We highlight that changing the local symmetry group of spacetime permits to construct a theory in which these two symmetries are combined into a putative gauge symmetry but with second-order field equations and non-trivial mass scales, unlike the original higher-order construction by Weyl. We prove that the gravitational field equations are equivalent to the (trace-free) Einstein field equations, ensuring their compatibility with known tests of general relativity. As a corollary, the effective cosmological constant is rendered radiatively stable due to Weyl invariance. A novel phenomenological consequence characteristic of this construction, potentially relevant for cosmological observations, is the existence of an energy scale below which effects associated with the non-integrability of spacetime distances, and an effective mass for the electromagnetic field, appear simultaneously (as dual manifestations of the use of Weyl connections). We explain how former criticisms against Weyl's ideas lose most of their power in its present reincarnation, which we refer to as Weyl relativity, as it represents a Weyl-invariant, unified description of both the Einstein and Maxwell field equations.
}
\end{abstract}

\maketitle

\tableofcontents

\section{Introduction}

One hundred years ago, the physical view of the world radically changed with the introduction of general relativity and its description of spacetime as an elastic continuum of geometric nature. At the dawn of general relativity, the electromagnetic field stood alone as something closer to gravity than to matter. No wonder, people started to think that the electromagnetic field may also be part of a geometric description of the world. In 1918 Weyl  proposed the first unified theory putting together gravity and electromagnetism~\cite{Weyl1918}. Although the theory was received with owe by Einstein himself, appreciating its beauty, he rapidly found what he considered a physical contradiction that invalidated the theory on physical grounds (see Einstein's appendix to Weyl's paper): the theory seems to predict a ``second clock effect'' which has not been observed (for a description of the second clock effect see, for instance, Chap. 19 in \cite{Penrose2004}). Since then, these developments have lived in a Platonic world of beautiful ideas that seem to have no place in the real world.\footnote{It is worth remarking that nonetheless Weyl's construction was the starting point of the extremely successful idea of gauge theory that now pervades all of theoretical physics \cite{Scholz2011}.
}

At present we can still argue that gravity and electromagnetism stand alone from the rest of ingredients of nature. There are other gauge fields, the mathematical description of which is formally close to that of electromagnetism. However, their effects are confined to the interstices of matter so that they do not have large distance effects: they do not pervade the universe as gravity and electromagnetism do, which makes for a great difference from a physical standpoint. Thus, we think that it still makes sense to wonder whether these two phenomena could be part of a single geometrical framework. Moreover, we wonder whether these two theories could be made formally more similar to one another, for instance by sharing the same symmetry: conformal invariance (recall that the Einstein-Maxwell theory, though somewhat surprisingly due to the massless character of the excitations described, is not conformally invariant). This paper revises Weyl's ideas and their criticisms at the light of recent new developments by the authors \cite{Barcelo2014,Carballo-Rubio2015a} that are ultimately related to the theory known as unimodular gravity, or its extension Weyl-transverse gravity\footnote{Weyl-transverse gravity is an extension of unimodular gravity that is manifestly invariant under Weyl transformations, and hence displays the same number of generators of local symmetries as general relativity. Unimodular gravity is the gauge-fixed version of Weyl-transverse gravity, and not general relativity as it is sometimes claimed.} \cite{Anderson1971,Unruh1988,Unruh1989,Henneaux1989,Ng1990,Ng1990b,Bombelli1991,Ng1999,Finkelstein2000,Smolin2009,Ellis2011,Ellis2013}; see \cite{Alvarezetal2006,Jain2012,Gao2014,Cho2014,Kluson2014,Kluson2014b,Bonifacio2015,Eichhorn2015,Basak2015,Nojiri2015,Bull2015,Nojiri2016,Nojiri2016b,Nojiri2016c,Oda2016,Oda2016b,Oda2016c,Bamba2016,Nassur2016,Saez-Gomez2016,Mori2017} for recent developments related either with this theory or the core physical ideas behind it. As far as we know our approach in this paper is completely different, in both its intentions and emphasis, from other attempts to resurrect Weyl's ideas such as, e.g.,~\cite{Salim1996,YuanHuang2013,Romero2012,Romero2015}.

It is convenient to clarify the terminology followed in this paper, as it deviates somewhat from customs in the literature. In the literature, conformal invariance and Weyl invariance are terms that are sometimes used interchangeably. However, in this paper these refer to different concepts. Conformal invariance will refer to the symmetry under local scale transformations of the gravitational and matter fields, in which these fields are multiplied by different powers (known as weights) of the same scale factor. This symmetry appears typically in diffeomorphism-invariant theories. On the other hand, Weyl invariance will refer to the symmetry under the combination of a local scale transformation of the gravitational field and a gauge transformation of the electromagnetic vector potential. That is, in the latter not only the gravitational field is multiplied by a scale factor, but also the electromagnetic vector potential is modified by the addition of a gradient that depends on the very same scale factor. Moreover, matter fields are not affected (that is, these are of weight zero), which is only possible in a theory in which the diffeomorphism group is reduced to its transverse subgroup with the introduction of a background (non-dynamical) volume form

\section{Weyl relativity}

\subsection{The Weyl connection and its curvature tensors}

In general relativity, the parallel transportation of vector orientations is non-integrable while the transportation of vector lengths is integrable. Weyl's proposal is that the latter quantity has to be also non-integrable through the introduction of  a new connection $\bar{\Gamma}^c_{ab}$ with an associated covariant derivative $\bar{\nabla}_a$ such that 
\begin{eqnarray}
\bar{\nabla}_cg_{ab}=\mathcal{Q}A_cg_{ab}.\label{eq:nocomp}
\end{eqnarray}
Here $\mathcal{Q}$ is a real constant that has in principle nothing to do with the electromagnetic charge (nor any of the charges that appear in the standard model of particle physics), but is a new charge associated with the non-integrability of vector lengths. The most general connection compatible with Eq. \eqref{eq:nocomp} can accommodate torsion, though it will be assumed in the following that this is not the case.\footnote{Also the most general connection in the Palatini approach may display an additional vector field \cite{Bernal2016}, though this is however different from the Weyl connection we deal with in this paper.} In the absence of torsion, one can deduce from Eq. \eqref{eq:nocomp} the components of the connection in a coordinate basis:
\begin{eqnarray}
\bar{\Gamma}^c_{ab} &=& {1 \over 2}g^{cd} 
\left[ (\partial_a - \mathcal{Q} A_a ) g_{bd} + (\partial_b - \mathcal{Q} A_b) g_{ad} - (\partial_d - \mathcal{Q} A_d) g_{ab} \right]
\nonumber\\
&=&
\Gamma^c_{ab}-{\mathcal{Q} \over 2}\left[\delta^c_a A_b + \delta^c_b A_a - A^{c} g_{ab} \right],\label{eq:wconn}
\end{eqnarray}
where $\Gamma^c_{ab}$ corresponds to the Levi-Civita connection. The Levi-Civita covariant derivative $\nabla_a$ satisfies the integrability condition
\begin{equation}
\nabla_cg_{ab} =0.
\end{equation}
To differentiate these two different structures we will call $\bar{\nabla}_a$ the Weyl covariant derivative, and $\bar{\Gamma}^c_{ab}$ the Weyl connection. 

The construction of the curvature tensors of the Weyl connection follows the same rules as for any other connection (see, e.g., the discussion in Wald's book \cite{Wald1984}). For instance, its Riemann curvature tensor is defined as a measure of the noncommutativity of the Weyl covariant derivative $\bar{\nabla}_a$, taking in a coordinate basis the form
\begin{equation}
\bar{R}^d_{\,\,abc}=\partial_b\bar{\Gamma}^d_{ac}-\partial_c\bar{\Gamma}^d_{ab}+\bar{\Gamma}^d_{be}\bar{\Gamma}^e_{ac}-
\bar{\Gamma}^d_{ce}\bar{\Gamma}^e_{ab}.
\end{equation}
The Riemann tensor of the Weyl connection displays the symmetries of the Riemann tensor of the Levi-Civita connection, and satisfies the two Bianchi identities due to the absence of torsion.

The corresponding Ricci tensor is given by
\begin{equation}
\bar{R}_{ab}=\bar{R}^c_{\,\,acb}=\partial_c\bar{\Gamma}^c_{ab}-\partial_b\bar{\Gamma}^c_{ac}+\bar{\Gamma}^c_{ce}\bar{\Gamma}^e_{ab}-
\bar{\Gamma}^c_{be}\bar{\Gamma}^e_{ac}.
\end{equation}
It is important to keep in mind that the Ricci tensor of the Weyl connection is not symmetric. It is useful to split the Ricci tensor in its symmetric and antisymmetric parts,
\begin{equation}
\bar{R}_{ab}=\bar{R}_{(ab)}+\bar{R}_{[ab]}.
\end{equation}
The respective explicit forms of these quantities are given by
\begin{equation}
\bar{R}_{(ab)}=\partial_c\bar{\Gamma}^c_{ab}-\partial_{(b}\bar{\Gamma}^c_{a)c}+\bar{\Gamma}^c_{ce}\bar{\Gamma}^e_{ab}-
\bar{\Gamma}^c_{be}\bar{\Gamma}^e_{ac},\label{eq:rictendef}
\end{equation}
and
\begin{equation}
\bar{R}_{[ab]}=-\partial_{[b}\bar{\Gamma}^c_{a]c}=\mathcal{Q}(\partial_bA_a-\partial_aA_b)=-\mathcal{Q}F_{ab}.
\end{equation}
The last equation introduces the electromagnetic field tensor $F_{ab}$ as the antisymmetric part of the Ricci tensor of the Weyl connection.

With these ingredients, two curvature invariants can be introduced. One of them is the Ricci scalar, defined as
\begin{equation}
\bar{R}=g^{ab}\bar{R}_{ab},\label{eq:cursca1}
\end{equation}
which only picks up contributions from symmetric part of the Ricci tensor $R_{(ab)}$. An additional curvature scalar, constructed using the antisymmetric part of the Ricci tensor $R_{[ab]}$, is given by
\begin{equation}
K=g^{ac}g^{bd}\bar{R}_{[ab]}\bar{R}_{[cd]}=\mathcal{Q}^2g^{ac}g^{bd}F_{ab}F_{cd},\label{eq:cursca2}
\end{equation}
which can be recognized as the combination used in covariant formulations of the Maxwell field equations. Both curvature scalars in Eqs. \eqref{eq:cursca1} and \eqref{eq:cursca2} will be used to construct the action of Weyl relativity.

An important property of the Weyl connection and its curvature tensors is their behavior under the simultaneous transformation
\begin{eqnarray}
g_{ab} \longrightarrow \Omega^2 g_{ab},\qquad\qquad A_a \longrightarrow A_a + {2 \over \mathcal{Q}} \partial_a \ln \Omega,\label{eq:wtrans}
\end{eqnarray}
that we will call Weyl transformation. From their definition, it can be checked that the Weyl connection is Weyl invariant. This follows from the behavior of the Levi-Civita connection under the transformations in Eq. \eqref{eq:wtrans}, namely
\begin{equation}
\Gamma^c_{ab}\longrightarrow \Gamma^c_{ab}+\delta^c_a\partial_b\ln\Omega+\delta^c_b\partial_a\ln\Omega-g_{ab}\partial^c\ln\Omega,\label{eq:gammatrans}
\end{equation}
and the transformation of the vector field $A_a$ in Eq. \eqref{eq:wtrans}. 

Note that Weyl transformations leave invariant the Weyl connection \eqref{eq:wconn}. These transformations also leave invariant the Riemann and Ricci tensors (both the symmetric and antisymmetric parts of the latter separately). On the other hand, none of the curvature scalars are invariant under Weyl transformations: the Ricci scalar verifies
\begin{equation}
\bar{R}\longrightarrow \Omega^{-2}\bar{R},\label{eq:rscaltrans}
\end{equation}
while the scalar quadratic in the antisymmetric part of the Ricci tensor behaves as
\begin{equation}
K\longrightarrow \Omega^{-4}K.\label{eq:rscaltrans2}
\end{equation}
%

\subsection{A second-order action principle \label{sec:weylact}}

In the previous section we have introduced the Weyl connection and the curvature tensors associated with it, discussing the transformation rules of these different quantities with respect to Weyl transformations. The latter description follows tightly the original construction by Weyl and later works. It is in this section where our discussion will deviate from the original discussion, following a different path even if starting from the very same setting.

The observation that prompts the discussion is that the action that would resemble more closely the Einstein-Hilbert action being at the same time Weyl invariant 
cannot be constructed only in terms of $g_{ab}$. This is due to the transformation rule \eqref{eq:rscaltrans}, which implies that
\begin{equation}
\sqrt{-g}\bar{R}\longrightarrow \Omega^{2}\sqrt{-g}\bar{R}.\label{eq:noninv}
\end{equation}
Therefore this scalar density cannot be used to construct a Weyl-invariant action through integration. 

As far as we are able to tell, we have come to the conclusion that there are two possible ways to sort this out (without considering additional fields):
\begin{itemize}
\item{\emph{The familiar, higher-derivative road:} Even if the Ricci scalar is not apt for constructing the action, other curvature invariants could be used, though these have to be of higher order in the derivatives of the gravitational field. The resulting theories are typically formulated by means of higher derivative Lagrangian densities, for instance
\begin{equation}
\sqrt{-g}\,g^{ac}g^{bd}\bar{R}_{ab}\bar{R}_{cd}.
\end{equation}
In many other ocassions people work with the Bach Lagrangian $\sqrt{-g} C_{abcd} C^{abcd}$ (where $C_{abcd}$ is the Weyl tensor of $g_{ab}$), which is directly Weyl invariant without having to use the Weyl connection; see for instance \cite{Mannheim1989,Sultana2010,Sultana2012,Sultana2012b,Wheeler2013,Kiefer2017}. This choice also leads to a higher-derivative theory. These higher derivative theories add typically new degrees of freedom to the gravitational sector and not all of them pass the stringent observational tests of general relativity~\cite{Edery1997,Flanagan2006,Cutajar2014}. From another side, current astronomical observations are telling us that a cosmological constant term is probable to be there (we say just probably because the observed cosmological acceleration might have other origin than a cosmological constant \cite{Tomita2000,Tomita2001,Fleury2013,Fleury2014}). So it would be desirable to have field equations able to accomodate a cosmological constant.}

\item{\emph{The unexplored, second-order road:} A loophole to the necessity of higher-derivative theories lies in the consideration of background (that is, non-dynamical) quantities. These quantities are ubiquitous in physics, being an unmistakable trace of the effective nature of the descriptions involving them. A simple example would be quantum electrodynamics, which uses a background Minkowskian structure; another example which is closer in spirit to the present discussion is Weyl-transverse gravity. For the purposes of this paper the necessary geometric structure is simpler than a complete metric structure, being just a volume form $\bm{\omega}$ in spacetime,\footnote{A metric structure is composed of two independent structures: volume and causal structures \cite{Ehlers2012,Hawking1973,Bradonjic2011}.} the introduction of which reduces the relevant symmetry group from diffeomorphisms to transverse diffeomorphisms. This volume form permits to construct a Lagrangian density while offering the possibility of clearing off the non-invariant terms in Eq. \eqref{eq:noninv}. The associated volume element will be denoted by $\sqrt{-\omega}$, being it a scalar density with the same weight as $\sqrt{-g}$. As we show in the following, this mild background structure alone permits to construct Weyl-invariant, second-order field equations.}
\end{itemize}

Using the background volume element, we can define the following Lagrangian density:
\begin{eqnarray}
\mathscr{L}_g=\frac{1}{2\kappa}\sqrt{-\omega}\,(g/\omega)^{1/4}\bar{R},
\label{eq:glagrangian}
\end{eqnarray}
where $\kappa=8\pi G/c^3$ with $G$ the gravitational constant and $c$ the speed of light. Let us emphasize two properties of the above Lagrangian density. The first one is that the background volume element permits to tune the power of the determinant of the gravitational field $g_{ab}$ in front of the Ricci scalar $\bar{R}$ to $(-g)^{1/4}$ instead of $(-g)^{1/2}$, while still leading to a scalar density that can be integrated. This guarantees that the outcome is Weyl invariant, unlike the combination in Eq. \eqref{eq:noninv}. In other words, it can be equivalently said that the minimum background structure needed to lead to a second-order, Weyl-invariant action for the gravitational field is a background volume form. From the perspective of symmetries, the construction will be no longer diffeomorphism invariant: the introduction of a background volume form breaks this symmetry group down to its transverse subgroup, defined as the set of diffeomorphisms that leave the volume form invariant.

The second property is that, as explained in the previous section, the Ricci scalar $\bar{R}$ has the same functional form as the equivalent quantity in Riemannian geometry, but using the Weyl connection \eqref{eq:wconn} in which the ordinary derivatives $\partial_a$ of the gravitational field are replaced by covariant derivatives $\partial_a-\mathcal{Q}A_a$ with respect to the vector field $A_a$. Therefore, while using a background volume element is certainly reminiscent of Weyl-transverse (or unimodular) gravity \cite{Carballo-Rubio2015a}, the scope of the present construction is much larger from both conceptual and practical perspectives. A clear formal difference lies in the internal structure of the Ricci scalar $\bar{R}$ used to define the action \eqref{eq:glagrangian}, which displays a genuine gauge invariance due to the introduction of the covariant derivative with respect to $A_a$. The very same feature leads to a different coupling between gravitational and electromagnetic fields, the consequences of which are explained in Secs. \ref{sec:feq} and \ref{sec:matt}. Indeed, it can be argued that Weyl relativity supersedes in some sense Weyl-transverse gravity as a broader construction: as it will be discussed, the latter is a proper subset of the present construction. This observation leads to refreshing suggestions about some of its most interesting properties.

While Eq. \eqref{eq:glagrangian} contains the allowed kinetic terms for the gravitational field $g_{ab}$, kinetic terms for the vector field $A_a$ should also be included. We will discuss the inclusion of matter fields later, in Sec. \ref{sec:matt}. Also there is an additional scalar curvature $K$, defined in Eq. \eqref{eq:cursca2}, which remains to be used. This obvious choice results in the standard Lagrangian density for the electromagnetic vector potential $A_a$:
\begin{equation}
\mathscr{L}_A=-\frac{1}{4\mu_0\mathcal{Q}^2}\sqrt{-g}\,K=-\frac{1}{4\mu_0}\sqrt{-g}\,g^{ac}g^{bd}F_{ab}F_{cd}.
\end{equation}
In this equation, the vacuum permeability $\mu_0$ has been introduced. This scalar density is Weyl invariant by itself, a feature which stems from the conformal invariance of the electromagnetic field in 4 dimensions.

This closes the construction of a theory of gravity and electromagnetism (with no matter fields) that materializes Weyl's insights, combining the gauge transformations of electromagnetism and local scale transformations of the gravitational field in the same symmetry. Let us then summarize the form of the action of Weyl relativity in the absence of matter:
\begin{equation}
\mathscr{S}=\int\text{d}^4x\left(\mathscr{L}_g+\mathscr{L}_A\right)=\int\text{d}^4x\sqrt{-\omega}\left[\frac{1}{2\kappa}(g/\omega)^{1/4}\bar{R}-\frac{1}{4\mu_0}(g/\omega)^{1/2}g^{ac}g^{bd}F_{ab}F_{cd}\right].\label{eq:totact1}
\end{equation}
Each of the two pieces of this action are separately invariant under Weyl transformations \eqref{eq:wtrans}, being constructed using the two curvature scalars \eqref{eq:cursca1} and \eqref{eq:cursca2} and the background volume form when necessary. The result is indeed the most general action using these ingredients only, and leading to second-order equations that are preserved under the required symmetries. This is to be compared with the situation in the Einstein-Maxwell theory in which the gravitational term, namely the Einstein-Hilbert action, breaks the symmetry under Weyl transformations. Note also that a cosmological constant term cannot be included in Eq. \eqref{eq:totact1} without breaking this symmetry; in Sec. \ref{sec:cc} we will explain briefly how an effective cosmological constant is nevertheless present in the field equations, and the non-trivial implications that follow.

\subsection{The field equations \label{sec:feq}}

Armed with the action \eqref{eq:totact1} it is straightforward to deduce the field equations. Let us nevertheless display this procedure in detail in order to highlight the main differences between using the gravitational action \eqref{eq:glagrangian} and the Einstein-Hilbert action.

Before entering into details, it is worth bearing in mind that the trace of the gravitational field equations does not have to be identically zero. While this must certainly be the case for a theory that is invariant under conformal transformations, invariance under Weyl transformations \eqref{eq:wtrans} leads to the milder identity to be satisfied by the action $\mathscr{S}$. A similar assertion can be done about the conservation of the source of the Maxwell field equations. Indeed, considering an infinitesimal Weyl transformation, namely making $\Omega=1+\epsilon$ in Eq. \eqref{eq:wtrans}, one has
\begin{equation}
\int\text{d}^4x\left(\frac{\delta\mathscr{L}}{\delta g^{ab}}\delta g^{ab}+\frac{\delta\mathscr{L}}{\delta A_a}\delta A_a\right)=2\int\text{d}^4x\left(-\frac{\delta\mathscr{L}}{\delta g^{ab}}\epsilon g^{ab}+\frac{1}{\mathcal{Q}}\frac{\delta\mathscr{L}}{\delta A_a}\partial_a\epsilon\right)=0.
\end{equation}
Integrating by parts one gets the relation
\begin{equation}
g^{ab}\frac{\delta\mathscr{L}}{\delta g^{ab}}+\frac{1}{\mathcal{Q}}\partial_a\frac{\delta\mathscr{L}}{\delta A_a}=0.\label{eq:newward}
\end{equation}
This equation permits to deduce that the gravitational field equations are not identically traceless, nor the electromagnetic equations identically conserved, but that both conditions are effectively verified on the space of solutions of the theory.

Let us start with the gravitational field equations, thus considering the variation of the action \eqref{eq:totact1} with respect to $g^{ab}$. The variation of the Ricci tensor is given in terms of the variations of the Christoffel symbols as
\begin{equation}
\delta \bar{R}_{ab}=\bar{\nabla}_c\delta\bar{\Gamma}^c_{ab}-\bar{\nabla}_a\delta\bar{\Gamma}^c_{cb}=\delta^{de}_{ac}\bar{\nabla}_e\delta\bar{\Gamma}^c_{db},
\end{equation}
where we have introduced the generalized Kronecker delta
\begin{equation}
\delta^{de}_{ac}=2\delta^{[d}_a\delta^{e]}_c.
\end{equation}
Taking into account that $\delta g=-g\,g_{ab}\delta g^{ab}$, one has then
\begin{equation}
2\kappa\,\delta\mathscr{L}_g=\sqrt{-\omega}\,(g/\omega)^{1/4}\left[\bar{R}_{(ab)}-\frac{1}{4}\bar{R}\,g_{ab}\right]\delta g^{ab}+\sqrt{-\omega}\,(g/\omega)^{1/4}g^{ab}\delta^{de}_{ac}\bar{\nabla}_e\delta\bar{\Gamma}^c_{db}.\label{eq:int1}
\end{equation}
The first term in the previous equation is proportional to the traceless part of the Ricci tensor. In the Einstein-Hilbert scenario, the last term in the previous equation is a boundary term, and hence irrelevant to the field equations. But here it leads to a non-trivial contribution to the field equations.

Indeed, as we show in App. \ref{sec:app}, the last term in Eq. \eqref{eq:int1} can be written up to a boundary term in the form
\begin{equation}
\sqrt{-g}\,(\omega/g)^{1/4}\delta g^{ab}\Delta_{ab},
\end{equation}
where
\begin{align}
\Delta_{ab}&=-(g/\omega)^{1/4}\left(\delta^e_{(a}\delta^f_{b)}-\frac{1}{4}g^{ef}g_{ab}\right)\bar{\nabla}_f\left[\partial_e(\omega/g)^{1/4}+\mathcal{Q}A_e\,(\omega/g)^{1/4}\right]\nonumber\\
&-\mathcal{Q}\left(A_{(a}\delta^c_{b)}-\frac{1}{4}A^cg_{ab}\right)\left[\partial_c\ln(\omega/g)^{1/4}+\mathcal{Q}A_c\right]\nonumber\\
&+\frac{3}{4}(g/\omega)^{1/4}g^{ef}g_{ab}\nabla_f\left[\partial_e(\omega/g)^{1/4}+\mathcal{Q}A_e\,(\omega/g)^{1/4}\right].\label{eq:findel}
\end{align}
It is worth checking that this expression is invariant under Weyl transformations. As written in Eq. \eqref{eq:findel}, the quantity $\Delta_{ab}$ is split in its traceless and pure trace parts. These two pieces have to be independently invariant under Weyl transformations \eqref{eq:wtrans}. It is easy to show that the trace part is invariant, being it given by
\begin{align}
&\frac{3}{4}(g/\omega)^{1/4}g^{ef}g_{ab}\nabla_f\left[\partial_e(\omega/g)^{1/4}+\mathcal{Q}A_e\,(\omega/g)^{1/4}\right]\nonumber\\
&=\frac{3}{4}g_{ab}\frac{1}{\sqrt{-g}}(g/\omega)^{1/4}\partial_e\left(\sqrt{-g}\,g^{ef}\left[\partial_b(\omega/g)^{1/4}+\mathcal{Q}A_f\,(\omega/g)^{1/4}\right]\right).\label{eq:tracefeq}
\end{align}
The last expression is obtained using the standard form of the divergence of a vector field in Riemannian geometry; it is then straigthforward to show that the trace part of $\Delta_{ab}$ is invariant under Weyl transformations \eqref{eq:wtrans}. 

Concerning the traceless part of $\Delta_{ab}$, namely the two first lines in Eq. \eqref{eq:findel}, it is given by the traceless part of the tensor
\begin{equation}
-(g/\omega)^{1/4}\bar{\nabla}_{(a}\left[\partial_{b)}(\omega/g)^{1/4}+\mathcal{Q}A_{b)}(\omega/g)^{1/4}\right]-\mathcal{Q}A_{(a}\left[\partial_{b)}\ln(\omega/g)^{1/4}+\mathcal{Q}A_{b)}\right],
\end{equation}
which is indeed invariant under Weyl transformations \eqref{eq:wtrans}.

Therefore, the following Weyl-invariant gravitational field equations are obtained, taking into account that the variation of the electromagnetic part of the action, or in other words the Lagrangian density $\mathscr{L}_A$, is the standard one:
\begin{equation}
\bar{R}_{(ab)}-\frac{1}{4}\bar{R}\,g_{ab}+\Delta_{ab}=\kappa\,(g/\omega)^{1/4}\,T^A_{ab},\label{eq:gravequs}
\end{equation}
where $\Delta_{ab}$  is given in Eq. \eqref{eq:findel} and $T^A_{ab}$ is the Maxwell stress-energy tensor,
\begin{eqnarray}
T^{A}_{ab}=g^{cd} F_{ac}F_{bd} -{1 \over 4} g_{ab} (g^{ef}g^{cd} F_{ec}F_{fd}),\label{eq:maxset1}
\end{eqnarray}
A number of facts about these equations are worth remarking. Most of the pieces in these gravitational field equations are easily identified: the trace-free part of the Ricci tensor, and the Maxwell stress-energy tensor. The multiplicative factor in front of $T^A_{ab}$ ensures that the right-hand side of the gravitational field equations is of zero conformal weight. Apart from the free electromagnetic field, we have to keep in mind that $\bar{R}_{ab}$ contains additional dependencies on the electromagnetic gauge field through the Weyl connection defined in Eq. \eqref{eq:wconn}. We will write explicitly the form of the corresponding terms at the end of this section.

Concerning the dynamics of the electromagnetic field, the electromagnetic field equations can be easily obtained using the previous discussion. There are no surprises in the contribution coming from the variation of the piece $\mathscr{L}_A$ of the Lagrangian density with respect to $A_a$, which is given up to boundary terms by
\begin{align}
\delta\mathscr{L}_A=-\frac{1}{\mu_0}\delta A_a\partial_b[\sqrt{-g}\,g^{ac}g^{bd}F_{cd}].
\end{align}
However, the variation coming from $\mathscr{L}_g$ must be also taken into account, namely
\begin{align}
\delta\mathscr{L}_g&=\frac{1}{2\kappa}\sqrt{-\omega}\,(g/\omega)^{1/4}g^{ab}\delta R_{ab}\nonumber\\
&=-\frac{1}{2\kappa}\sqrt{-g}\,\delta^{de}_{ac}\delta\bar{\Gamma}^c_{db}\left(\nabla_e\left[(\omega/g)^{1/4}g^{ab}\right]+\mathcal{Q}A_e\,(\omega/g)^{1/4}g^{ab}\right),
\end{align}
where now the variations $\delta\bar{\Gamma}^c_{eb}$ with respect to $A_a$ are to be considered (see App. \ref{sec:app} for the details). The electromagnetic field equations are then given by
\begin{equation}
\nabla_bF^{ab}+\frac{3\mu_0\mathcal{Q}}{2\kappa}(\omega/g)^{1/4}g^{ab}\left[\partial_b\ln(\omega/g)^{1/4}+\mathcal{Q}A_b\right]=0.\label{eq:eleequs}
\end{equation}
It is not difficult to check that these equations are invariant under Weyl transformations \eqref{eq:wtrans}, using the standard relation for the divergence of an antisymmetric tensor:
\begin{equation}
\nabla_bF^{ab}=\frac{1}{\sqrt{-g}}\partial_b[\sqrt{-g}\,g^{ac}g^{bd}F_{cd}].
\end{equation}
Note that Eq. \eqref{eq:eleequs} implies that the dimensionful combination $\mu_0\mathcal{Q}/\kappa$ should have dimensions of the vector potential $A_a$ times inverse of length.

The field equations \eqref{eq:gravequs} and \eqref{eq:eleequs} can be partially solved in order to illustrate some of its most interesting properties. As it was advanced in Eq. \eqref{eq:newward}, the gravitational field equations \eqref{eq:gravequs} are not traceless, nor the electromagnetic field equations \eqref{eq:eleequs} conserved, being these two features intertwined. The trace of the gravitational field equations is given, as advanced in Eq. \eqref{eq:tracefeq}, by
\begin{equation}
g^{ab}\Delta_{ab}=\frac{3}{\sqrt{-g}}(g/\omega)^{1/4}\partial_e\left(\sqrt{-g}\,g^{ef}\left[\partial_b(\omega/g)^{1/4}+\mathcal{Q}A_f\,(\omega/g)^{1/4}\right]\right),\label{eq:tracecond0}
\end{equation}
which in the absence of matter fields (or if coupled to traceless stress-energy tensors, which as shown below it is indeed the case) must be zero. Then, the general solution to this equation can be written explicitly, and corresponds to a specific Helmholtz decomposition of the vector field $A_a$ in which its longitudinal part is completely determined by the gravitational field:
\begin{equation}
A_a=-\frac{1}{Q}\partial_a\ln(\omega/g)^{1/4}+\hat{A}_a,\label{eq:tracecond}
\end{equation}
where $\hat{A}_a$ is a vector field satisfying
\begin{equation}
\nabla_a\left[(\omega/g)^{1/4}\hat{A}^a\right]=\frac{1}{\sqrt{-g}}\partial_a\left[\sqrt{-g}\,g^{ab}(\omega/g)^{1/4}\hat{A}_b\right]=0.\label{eq:divcond}
\end{equation}
This equation is invariant under Weyl transformations, noticing that from its very definition in Eq. \eqref{eq:tracecond}, $\hat{A}_a$ transforms trivially, namely $\hat{A}_a\longrightarrow \hat{A}_a$.

Eq. \eqref{eq:tracecond} permits to simplify the form of the field equations. For instance, the Weyl connection is decomposed under Eq. \eqref{eq:tracecond} as
\begin{align}
\bar{\Gamma}^c_{ab}&= {1 \over 2}\hat{g}^{cd} 
\left[\partial_a\hat{g}_{bd} + \partial_b\hat{g}_{ad}-\partial_d\hat{g}_{ab}\right]-{\mathcal{Q} \over 2}\left[\delta^c_a \hat{A}_b + \delta^c_b \hat{A}_a - \hat{A}^{c} g_{ab} \right]\nonumber\\
&=\hat{\Gamma}^c_{ab}-{\mathcal{Q} \over 2}\left[\delta^c_a \hat{A}_b + \delta^c_b \hat{A}_a - \hat{A}^{c} g_{ab} \right],\label{eq:gammahat}
\end{align}
where the following composite, Weyl-invariant gravitational field has been defined:
\begin{equation}
\hat{g}_{ab}=(\omega/g)^{1/4}g_{ab}.\label{eq:ghat1}
\end{equation}
Eq. \eqref{eq:gammahat} permits to write down a similar decomposition for curvature tensors \cite{YuanHuang2013}; e.g.,
\begin{equation}
\bar{R}_{ab}-\frac{1}{4}\bar{R}g_{ab}=\hat{R}_{ab}-\frac{1}{4}\hat{R}\hat{g}_{ab}+\mathcal{Q}\left(\hat{\nabla}_{(a}\hat{A}_{b)}-\frac{1}{4}g_{ab}\hat{\nabla}_c\hat{A}^c\right)+\frac{\mathcal{Q}^2}{2}\left(\hat{A}_a\hat{A}_b-\frac{1}{4}g_{ab}\hat{A}^c\hat{A}_c\right),\label{eq:riccidec}
\end{equation}
where $\hat{\nabla}$ is the covariant derivative associated with $\hat{\Gamma}^c_{ab}$ as defined in Eq. \eqref{eq:gammahat}. Also Eq. \eqref{eq:findel} is simplified down to
\begin{align}
\Delta_{ab}=-\mathcal{Q}\left(\hat{\nabla}_{(a}\hat{A}_{b)}-\frac{1}{4}g_{ab}\hat{\nabla}_c\hat{A}^c\right)-2\mathcal{Q}^2\left(\hat{A}_a\hat{A}_b-\frac{1}{4}g_{ab}\hat{A}_c\hat{A}^c\right)
+{3 \over 4}{\cal Q}\hat{g}_{ab} \hat{\nabla}_c \hat{A}^c.
\label{eq:deltadec}
\end{align}
Overall, the trace of the gravitational field equations can be solved independently in this way. This leaves only the traceless part of the gravitational field equations in Eq. \eqref{eq:gravequs} which, using Eqs. \eqref{eq:riccidec} and \eqref{eq:deltadec} can be written as
\begin{equation}
\hat{R}_{ab}-\frac{1}{4}\hat{R}\hat{g}_{ab}-\frac{3\mathcal{Q}^2}{2}\left(\hat{A}_a\hat{A}_b-\frac{1}{4}g_{ab}\hat{A}^c\hat{A}_c\right)=\kappa\,(g/\omega)^{1/4}\,T^{A}_{ab},\label{eq:zfin}
\end{equation}
or equivalently,
\begin{equation}
\hat{R}_{ab}-\frac{1}{4}\hat{R}\hat{g}_{ab}=\kappa\left(\hat{T}^{\hat{A}}_{ab}-\frac{1}{4}\hat{T}^{\hat{A}}_{cd}\hat{g}^{cd}\hat{g}_{ab}\right),\label{eq:zfin1}
\end{equation}
where $\hat{T}^{\hat{A}}_{ab}$ is the stress-energy tensor of a Proca field with mas proportional to $\mathcal{Q}^2$ and written in terms of $\hat{g}_{ab}$:
\begin{align}
&\hat{T}^{\hat{A}}_{ab}-\frac{1}{4}\hat{T}^{\hat{A}}_{cd}\hat{g}^{cd}\hat{g}_{ab}\nonumber\\
&=\hat{g}^{cd} F_{ac}F_{bd} -{1 \over 4} \hat{g}_{ab} (\hat{g}^{ef}\hat{g}^{cd} F_{ec}F_{fd})+\frac{3\mathcal{Q}^2}{2\kappa}\left(\hat{A}_a\hat{A}_b-\frac{1}{4}\hat{g}_{ab}\hat{g}^{cd}\hat{A}_c\hat{A}_d\right).
\end{align}
Accordingly, the electromagnetic field equations take the form associated with a Proca field,
\begin{equation}
\nabla_bF^{ab}+\frac{3\mu_0\mathcal{Q}^2}{2\kappa}(\omega/g)^{1/4}g^{ab}\hat{A}_b=0.\label{eq:finmax}
\end{equation}
Note that in the left-hand side of the gravitational field equations \eqref{eq:zfin1}, curvature tensors are evaluated on the composite gravitational field defined in \eqref{eq:ghat1}, which makes these equations invariant under Weyl transformations (recall that $\hat{A}_a$ remains unchanged under these transformations). In these equations, $F_{ab}$ is indistinctly the field strength associated with $A_a$ or $\hat{A}_a$. Eqs. \eqref{eq:zfin1} and \eqref{eq:finmax} are the field equations in the absence of additional matter fields.

Summarizing, the field equations of Weyl relativity in the absence of matter fields are equivalent to the trace-free Einstein equations characteristic of Weyl-transverse gravity, with a massive vector field. The additional polarization in the electromagnetic field just describes the additional degree of freedom in the theory besides the four degrees of freedom associated with massless gravitons and photons. This additional degree of freedom can be explicitly embedded in either the gravitational or the electromagnetic part of the theory, by using Weyl transformations (see the related discussion in Sec. \ref{sec:weyldiff}). It is quite interesting that this effective mass term arises due to the introduction on non-integrability effects on spacetime (both effects are controlled by the same dimensional constant $\mathcal{Q}$). In the limit of small field and small mass these equations become locally indistinguishable from standard electromagnetism.

\subsection{Introducing matter fields \label{sec:matt}}

The discussion has covered so far the construction of a unified theory of gravity and electromagnetism only, with no matter fields. Among other features, in the case of matter fields it would be necessary to introduce non-trivial mass scales. While in the original construction by Weyl it is not easy to include massive fields, it is interesting to note that the same ingredient which permits the construction of second-order field equations also makes room for the inclusion of non-trivial mass scales; a detailed discussion can be read in \cite{Carballo-Rubio2015a}.

To construct the matter action let us take a modified minimal prescription in the transition from flat to curved spacetimes. In this prescription, any flat metric $\eta_{ab}$ is replaced by $\hat{g}_{ab}$ defined in Eq. \eqref{eq:ghat1}. For fermion fields a similar definition exists for the vierbein field \cite{Carballo-Rubio2015a}. For instance, the action for a free massive scalar field would be given by
\begin{equation}
\mathscr{S}_\phi=\frac{1}{2}\int\text{d}^4x\sqrt{-\omega}\left(\hat{g}^{ab}\partial_a\phi\partial_b\phi-m^2\phi^2\right).
\end{equation}
Note that a general potential term $V(\phi)$ could be included instead of the mass term. The introduction of a background volume form $\bm{\omega}$ permits to write down a non-trivial mass term without spoiling the Weyl invariance of the theory; of course, this trades longitudinal diffeomorphisms (i.e., the diffeomorphisms that do not preserve the volume form $\bm{\omega}$) for Weyl transformations as the members of the symmetry group of the theory. 

On the other hand, the coupling of the electromagnetic field to matter fields is done through the standard prescription, replacing the derivatives in the kinetic terms of charged fields by covariant derivatives with respect to the gauge connection $A_a$. Following these specific minimal coupling prescriptions, fields with arbitrary spin and potential terms (and in particular, the entire standard model of particle physics) can be coupled both to gravitational and electromagnetic fields.

Let us now discuss the effect of a matter action $\mathscr{S}_M[g]=\mathscr{S}^0_M[\hat{g}(g)]$ composed of an arbitrary number of fields of different spin and mass. Let us define $\mathscr{S}_M^0[g]$ as the action in general relativity that is obtained following the standard minimal coupling prescription in the framework of the Einstein-Maxwell theory. Then one has
\begin{align}
\frac{\delta\mathscr{S}_M}{\delta g^{ab}}=\frac{\delta\mathscr{S}^0_M}{\delta \hat{g}^{cd}}\frac{\delta\hat{g}^{cd}}{\delta g^{ab}}&=(g/\omega)^{1/4}\frac{\delta\mathscr{S}^0_M}{\delta \hat{g}^{cd}}\left(\delta^c_{(a}\delta^d_{b)}-\frac{1}{4}g^{cd}g_{ab}\right)\nonumber\\
&=(g/\omega)^{1/4}\left(\frac{\delta\mathscr{S}^0_M}{\delta \hat{g}^{ab}}-\frac{1}{4}g^{cd}\frac{\delta\mathscr{S}^0_M}{\delta \hat{g}^{cd}}g_{ab}\right).
\end{align}
The latter equation implies that the difference between the minimal prescription that we follow here and the standard one is that variations with respect to $g_{ab}$ pick up only traceless tensors. In particular, this implies that the matter source of the gravitational field equations is given by the traceless part of the stress-energy tensor. Hence the trace of the gravitational field equations can be solved independently as in the vacuum case, leading to
\begin{equation}
\hat{R}_{ab}-\frac{1}{4}\hat{R}\hat{g}_{ab}=\kappa\left(\hat{T}_{ab}-\frac{1}{4}\hat{T}_{cd}\hat{g}^{cd}\hat{g}_{ab}\right),\label{eq:gravimat}
\end{equation}
where $\hat{T}_{ab}=\hat{T}^{\hat{A}}_{ab}+\hat{T}^M_{ab}$. On the other hand, the electromagnetic field equations take the form
\begin{equation}
\nabla_bF^{ab}+\frac{3\mu_0\mathcal{Q}^2}{2\kappa}(\omega/g)^{1/4}g^{ab}\hat{A}_b=J^a=\frac{\delta\mathscr{S}_M}{\delta A_a},\label{eq:elecrtmat}
\end{equation}
where $J^a$ is the standard electromagnetic current. Eqs. \eqref{eq:gravimat} and \eqref{eq:elecrtmat} are the field equations of Weyl relativity in the presence of an arbitrary matter content; for instance, $\hat{T}^M_{ab}$ may correspond to the stress-energy tensor of a perfect fluid.

\subsection{Weyl relativity and diffeomorphism invariance \label{sec:weyldiff}}

Up to here, our line of reasoning in constructing a Weyl-relativity theory has been heuristic and inspired by the presence of a fiduciary volume elements in Weyl-transverse gravity. However, there is an alternative logic to build up a theory that shares several important elements with the previous construction (although not all of them).

Let us for instance start from the two physical fields, $\{g_{ab}, A_a\}$, considered in Weyl's geometrical construction. One could ask whether it is possible to build a set of tensorial, second-order and Weyl-invariant field equations using only these two fields. The answer is no: without employing additional non-dynamical structures, the tensorial requirement turns up to be equivalent to requiring the diffeomorphism invariance of the construction. The obstruction to find such a construction is the same we discussed above regarding the different conformal weights of gravitational and electromagnetic sectors. The minimal way in which one can avoid this obstruction is by introducing a single additional scalar field $\phi$ that under 
Weyl transformations changes as
\begin{eqnarray}
\phi \to \phi + 2\ln \Omega.
\label{eq:scalarWtrans}
\end{eqnarray}
Then, one can construct a Lagrangian density of the form
\begin{eqnarray}
\mathscr{L}&=& {1 \over 2\kappa} \sqrt{-g} e^{-\phi} g^{ab}\bar{R}_{ab}
-{1 \over 4c \mu_0} \sqrt{-g} F_{ab} F_{cd} g^{ac} g^{bd}.
\\
&&
+{1 \over 2} \sqrt{-g} e^{-\phi} g^{ab} \left(\partial_a \phi - {\cal Q} A_a \right)\left(\partial_b \phi - {\cal Q} A_b \right)
- {1 \over 2\kappa} \sqrt{-g} \lambda e^{-2\phi}.
\nonumber
\label{eq:Wdiff}
\end{eqnarray}
It is not difficult to see that this theory is invariant under both diffeomorphisms and Weyl transformations. The theory has $10+4+1$ degrees of freedom (DOFs) and $5$ gauge symmetries. A rapid counting of degrees of freedom indicates that the gauge symmetries eliminate $10$ degrees of freedom, leaving a total of 5 physical degrees of freedom. Thus, apart from the $2 +2$ degrees of freedom associated with massless gravitons plus massless photons, we have one additional degree of freedom which can be associated to a mass for the photon or to an additional scalar field of gravitational origin. See \cite{Smolin1979,deCesare2016} for additional discussions on this theory as well as extensions to include additional standard model fields.

This diffeomorphism invariant theory is formally equivalent to the Stueckelberg model for an Abelian vector field (see, e.g., \cite{Ruegg2003} and references therein).
For instance, if one absorbs the scalar field into the definition of a new vector field $\tilde{A}_a \equiv A_a - (1/{\cal Q})\partial_b \phi$
and a new metric $\tilde{g}_{ab} \equiv e^{-\phi}g_{ab}$. The Lagrangian density is now 
\begin{eqnarray}
\mathscr{L}&&= {1 \over 2\kappa} \sqrt{-\tilde{g}} \tilde{g}^{ab}\bar{R}_{ab}(\tilde{g},\tilde{A})
-{1 \over 4c \mu_0} \sqrt{-\tilde{g}} \tilde{F}_{ab} \tilde{F}_{cd}\tilde{g}^{ac}\tilde{g}^{bd}.
\\
&&
+{1 \over 2} {\cal Q}^2 \sqrt{-\tilde{g}} \tilde{g}^{ab} \tilde{A}_a \tilde{A}_b
- {1 \over 2\kappa} \sqrt{-\tilde{g}} \lambda,
\nonumber
\label{eq:Wdiff-new}
\end{eqnarray}
where in the definition of the Weyl connection used in $\bar{R}_{ab}(\tilde{g},\tilde{A})$, the metric $g_{ab}$ is replaced by $G_{ab}$, and the connection $A_a$ by $\tilde{A}_a$. Removing a total divergence, this Lagrangian density can be rewritten as 
\begin{eqnarray}
\mathscr{L}&&= {1 \over 2\kappa} \sqrt{-\tilde{g}} \tilde{g}^{ab}R_{ab}(\tilde{g})
-{1 \over 4c \mu_0} \sqrt{-\tilde{g}} \tilde{F}_{ab} \tilde{F}_{cd} \tilde{g}^{ac} \tilde{g}^{bd}.
\\
&&
+\left({{\cal Q}^2 \over 2}-{3 \over 4\kappa} \right) \sqrt{-\tilde{g}} \tilde{g}^{ab} \tilde{A}_a \tilde{A}_b
- {1 \over 2\kappa} \sqrt{-\tilde{g}} \lambda,
\nonumber
\label{eq:Wdiff-newbis}
\end{eqnarray}
Written in this way, it is evident that the theory is equivalent to general relativity with a cosmological constant term coupled to a Proca field. Depending on the perspective adopted one could say that in the new variables Weyl invariance is hidden, or that the new theory does not display any Weyl invariance (the same observation applies to the Stueckelberg model).

Starting from the theory defined in Eq. \eqref{eq:Wdiff} there is an alternative way of eliminating the scalar field but now maintaining explicitly Weyl invariance, while breaking part of the diffeomorphism group. This can be done by introducing precisely a fiduciary volume element $\sqrt{-\omega}$. Then, one can tie the value of $\phi$ to that of the metric by setting
\begin{eqnarray}
\phi=\ln(g /\omega)^{1/4}.
\label{eq:scalarWtrans}
\end{eqnarray}
Defining now 
\begin{eqnarray}
\hat{g}_{ab}= (g /\omega)^{1/4} g_{ab},\qquad
\hat{A}_a=A_a- {1 \over {\cal Q}}\partial_a \ln(g /\omega)^{1/4},
\label{eq:scalarWtrans1}
\end{eqnarray}
we end up with the very Lagrangian density that we have described in Sec. \ref{sec:weylact}:
\begin{eqnarray}
\mathscr{L}&&= {1 \over 2\kappa} \sqrt{-\hat{g}} \hat{g}^{ab}\hat{R}_{ab}
-{1 \over 4c \mu_0} \sqrt{-\hat{g}} \tilde{F}_{ab} \tilde{F}_{cd} \hat{g}^{ac} \hat{g}^{bd}.
\\
&&
+\left({{\cal Q}^2 \over 2}-{3 \over 4\kappa} \right) \sqrt{-\hat{g}} \hat{g}^{ab} \hat{A}_a \hat{A}_b
- {1 \over 2\kappa} \sqrt{-\omega} \lambda.
\nonumber
\label{eq:Wdiff-newbis}
\end{eqnarray}
As discussed the theory is explicitly Weyl invariant but now it is not completely diffeormorphism invariant. This is so because the theory has an additional volume element: only the subset of diffeomorphism that leave this volume element un changed leave the form of the equations unchanged, that is, the $\xi^a$ such that $\hat{\nabla}_a \xi^a=0$. 

Constraining the scalar field $\phi$ to be given by Eq. \eqref{eq:scalarWtrans} at the level of the action (that is, before extracting the field equations) is not inconsequential. Due to this constraint, the variation of the action leads to the traceless Einstein equations instead of the standard Einstein field equations. This produces a subtle but very relevant change in some of the properties of the theory: the theory can incorporate an effective cosmological constant (given by an integration constant) without any problems of naturalness, due to the protection granted by Weyl invariance. This is explained in more detail in Sec. \ref{sec:cc}.

Starting from a theory that is both invariant under Weyl transformations and diffeomorphisms, it might seem unnatural to consider the restriction of the group of diffeomorphisms to its transverse part (defined by means of a given volume form). However, if one starts constructing a non-linear theory of gravity starting from a linear theory (Fierz-Pauli theory) defined in a fixed background, this possibility is certainly natural (and quite interesting) \cite{Barcelo2014}, as in this way one would retain some reminiscence, namely the volume element, of its origin. In any case, it is interesting to keep in mind the two alternative ways above to deal with a Weyl-invariant theory.

\section{Further observations}

\subsection{Regarding previous criticisms to Weyl's ideas}

The main criticism against Weyl's original theory is that it seems to predict the existence of a ``second clock'' effect:
two synchronized clocks initially at the same position but travelling through different regions, encountering different values of electromagnetic fields, should display different clock rates after being reunited (see, e.g., \cite{Avalos2016} for a recent discussion). Putting it in an even stronger way, two identical particles with identical masses, after being separated and reunited again should have different masses (mass is nothing but a length scale in this context). These two effects would be consequences of the non-integrability of spacetime
distances in Weyl-like theories. 

In more detail, non-integrability implies that distances (proper times) along a single geodesic do not have a physical meaning, not being Weyl invariant. However, if one compares the ratio between the distance between two events $P$ and $Q$ that are defined following two different curves $\gamma_1$ and $\gamma_2$, the result is physical and given by
\begin{eqnarray}
\frac{\tau_2}{\tau_1}=\exp{\left(\mathcal{Q}\int_{\gamma_2}A_a\,\text{d}l^a-\mathcal{Q}\int_{\gamma_1}A_a\,\text{d}l^a \right)}=\exp{\left(\mathcal{Q}\int_{\gamma_2}\hat{A}_a\,\text{d}l^a-\mathcal{Q}\int_{\gamma_1}\hat{A}_a\,\text{d}l^a \right)}.
\end{eqnarray}
Note that Weyl transformations leave the quantity above invariant, as
\begin{equation}
\int_{\gamma_2}\left(A_a+\frac{2}{\mathcal{Q}}\partial_a\ln\Omega\right)-\int_{\gamma_1}\left(A_a+\frac{2}{\mathcal{Q}}\partial_a\ln\Omega\right)=\int_{\gamma_2}A_a-\int_{\gamma_1}A_a.
\end{equation}
Hence in this theory only the quotient between distances in spacetime is well defined.
 
These non-integrability effect on spacetime distances have not been observed, which means that it should be possible to constrain the value of $\mathcal{Q}$ (note that for $\mathcal{Q}=0$ there is no second clock effect). In principle, experiments can just set upper bounds to its value. Interestingly, in the realization of Weyl's ideas presented in this paper, the parameter $\mathcal{Q}$ that controls non-integrability effects also leads to an effective mass for the electromagnetic field, as it can be read from the electromagnetic field equations \eqref{eq:elecrtmat}. We know from experiments that this effective mass has to be extremely small, implying at the same time the smallness of non-integrability effects. Note that it was claimed that the second clock effect should accumulate in time making it observable in current experiments, provided it existed~\cite{Pauli1958}. To our knowledge, there is no in-depth analysis of experiments that can be used to measure this effect, describing also how tight the corresponding bounds on the value of $\mathcal{Q}$ are. While it is clear that the second clock effect would introduce, for instance, an extra width in the spectral lines, whether or not this width can be indirectly detected in experimental setups not specifically designed to amplify and observe the effect itself may be quite non-trivial and, in our opinion, worth studying in detail. 

A second, and stronger comment is due to Weyl himself. He stressed that the notion of proper time that is constructed using the spacetime metric might not directly correspond to the proper time measured by material measuring devices. As Pauli remarked~\cite{Pauli1958}, if this view is correct there will not be ``direct conflicts with experiments''. However, the presence of fiduciary elements in the theory without direct operational meaning would seem to make the theory ``to have been robbed of its intrinsic convincing power''. Weyl's view is reinforced in the present construction by looking at the matter sector. The presence of rest masses in this sector does not break the Weyl invariance of the theory (recall that this Weyl invariance is different from other forms of conformal invariance in which the matter fields also acquire conformal weights). These masses are not affected whatsoever by the dynamics of the gravitational and electromagnetic fields. In a sense it is as if gravity plus electromagnetism were conformally invariant and non-conformal matter was breaking effectively this symmetry. Regarding the ``decrease of convincing power'' we assume here the view that these fiduciary elements might not have any role at the classical level but might acquire an active role in a higher-level description of reality. They are maybe reminding us about the effective nature of the theories at hand, whispering paths towards deeper descriptions.

\subsection{The cosmological constant and the quantum vacuum \label{sec:cc}}

We have explained in Sec. \ref{sec:matt} that the gravitational field equations take the form of the trace-free Einstein field equations:
\begin{equation}
\hat{R}_{ab}-\frac{1}{4}\hat{R}\hat{g}_{ab}=\kappa\left(\hat{T}_{ab}-\frac{1}{4}\hat{T}_{cd}\hat{g}^{cd}\hat{g}_{ab}\right).\label{eq:gravimat1}
\end{equation}
These represent a set of 9 equations instead of the 10 equations of general relativity. However, as is well known (see, e.g.,~\cite{Ellis2011,Ellis2013}), an exact equivalence with general relativity can be obtained by imposing as an additional requirement the conservation of the standard stress-energy tensor of the matter fields. This is an additional equation, that cannot be deduced from the symmetries of the theory (the symmetries of the theory only imply that the divergence of the stress-energy tensor is given by the gradient of an unknown function; it is possible that including this kind of non-conservation may be interesting phenomenologically \cite{Josset2016}):
\begin{equation}
\hat{\nabla}^a\hat{T}_{ab}=0.
\end{equation}
Under this condition, the entire set of Einstein equations is recovered,
\begin{eqnarray}
\hat{R}_{ab} - \frac{1}{2}\hat{R}\hat{g}_{ab} + \Lambda \hat{g}_{ab} =\kappa\,\hat{T}_{ab},\label{eq:gravimat2}
\end{eqnarray}
with a cosmological constant term appearing as an integration constant. That an effective cosmological constant arises in the gravitational field equations as an integration constant without breaking the Weyl symmetry (note that Eqs. \eqref{eq:gravimat2} are explicitly invariant under Weyl transformations) that forbid a cosmological term in the action is worth remarking. 

It is important to emphasize that the matter stress-energy in the right-hand side of Eq. \eqref{eq:gravimat2} accommodates general potential terms involving even non-trivial mass scales. This is due to the fact that only the complete stress-energy tensor, including potential terms, is conserved. Hence, it does not appear possible to phenomenologically distinguish Weyl relativity from general relativity coupled to matter at the classical level (if conservation of the matter stress-energy tensor is assumed).

Most importantly, the protection granted by Weyl symmetry forbids radiative corrections to the cosmological constant in Eq. \eqref{eq:gravimat2} or, in other words, makes the quantum vacuum non-gravitating. That is, there is no term in the Lagrangian density that is Weyl invariant, while at the same time leading to a cosmological constant in the field equations. This is preserved by quantum corrections to the theory, due to the fact that Weyl symmetry is not anomalous. Indeed, the same result that was shown in \cite{Carballo-Rubio2015a} can be generalized straightforwardly to Weyl relativity, as the main necessary element in the proof is the background volume element $\sqrt{-\omega}$. Moreover, in Weyl relativity it is made explicit that this protection is granted due to a genuine gauge symmetry involving the gauge field $A_a$, Weyl invariance. 

\subsection{Why Weyl relativity?\label{sec:why}}

General relativity follows from implementing the group of diffeomorphisms (Diff) as a gauge symmetry. Maintaining the number of degrees of freedom, it has been realized that an alternative group that can be used is the combination of transverse diffeomorphisms plus Weyl transformations (sometimes called WTDiff). Diff and WTDiff are distinct groups which are not contained in each other \cite{Alvarezetal2006}. We are used to defend the diffeomorphism invariance of Einstein relativity by arguing that physics should not depend on the coordinate system that an observer selects to describe the system. For the sake of the following discussion, let us exchange the previous words ``on the coordinate system'' by the more vague ``on the manner''. Starting from this second phrasing, one would have to find a precise definition of what does one mean by ``on the manner''. The WTDiff group offers an alternative definition to the traditional Diff invariance. This is why we have decided to associate the WTDiff group with a Weyl relativity principle and denote the resulting theory Weyl relativity.

In this regard, it is interesting to realize that the very operational construction of referential systems is tied up to the assumed theoretical framework. If one accepts that nature is Diff invariant, then it is possible to construct physical coordinate systems, for example, by sending light signals to four referential worldlines. Instead, if nature were WTDiff invariant, one would not be able to construct complete coordinate systems: one would have to construct coordinates modulo a (local) scale plus a separate scale retrieved by comparison with some fiducial one. So the WTdiff construction appears as self-consistent as the standard one based on Diff.  

As an example one can note that in Einstein relativity, it is allowed to stretch locally the physical description in just one direction by using a pure volume Diff transformation (take for instance $\nabla_a \xi^a=\rho $ with $\rho$ representing a string-like source). Effectively, a transformation in which this stretching becomes divergently large is as if locally there was an effective dimensional reduction: one dimension is divergently large with respect to the others. Should this type of transformation be part of the set of possible descriptions? We do not know.

These are precisely the type of transformations that do not belong to WTDiff: in Weyl relativity one can just change the shape of an object, respecting its volume (these correspond to TDiff transformations, which are part of both Diff and WTDiff), and then locally rescale the volume; locally any ``stretching'' of the physical description has to be isotropic. As a general lesson one can expect that in mesoscopic situations (meaning far from a global cosmological setting and far from a quantum microscopic regime) Diff and WTDiff should lead to equivalent phenomenologies, as it seems to be the case. However, when trying to elucidate the microscopic foundation of geometry or the cosmological behaviour of spacetime, they might lead to different pictures. 

\section{Summary}

Let us summarize the main properties of the theory of Weyl relativity introduced in this paper:
\begin{itemize}
\item{The field equations are quadratic in the derivatives of the gravitational, electromagnetic, and matter fields. These are invariant under Weyl transformations, a combination of local scale transformations of the gravitational fields and electromagnetic gauge transformations, that do not affect matter fields.}
\item{The field equations are closer to those of general relativity than in the original Weyl theory, being equivalent to the trace-free Einstein field equations that appear in Weyl-transverse gravity (or its gauge-fixed version, unimodular gravity) which, in vacuum or under the conservation of the matter stess-energy tensor in the presence of matter fields, are in turn equivalent to the entire set of Einstein field equations.}
\item{The non-linear interactions between the gravitational and electromagnetic fields manifest through an effective mass term for the electromagnetic vector potential. This mass scale is controlled by the dimensional constant that is also associated with the non-integrablity of spacetime distances, a characteristic phenomenon of using a Weyl connection. Phenomenological implications of these two features can be tested over long enough distances, most probably in cosmological observations.}
\item{Matter fields are not necessarily massless, but non-trivial mass scales can be included. Given that Weyl invariance does not affect matter fields, mass terms are compatible with the gauge symmetries of the theory. Equivalently, this is possible due to the introduction of a background (non-dynamical) volume form.}
\item{Weyl invariance forbids the inclusion of the cosmological constant term in the action. Nevertheless, the theory accommodates a cosmological constant arising as an integration constant in the gravitational field equations. Thanks to this subtle mechanism the field equations allow solutions with non-zero cosmological constant, ensuring at the same time that this quantity is protected by the gauge symmetries of the theory from acquiring a large value due to the fluctuations of the quantum vacuum.}
\end{itemize}

 
\acknowledgments
Financial support was provided by the Spanish MINECO through the projects FIS2011-30145-C03-01, FIS2011-30145-C03-02, FIS2014-54800-C2-1, FIS2014-54800-C2-2 (with FEDER contribution), and by the Junta de Andaluc\'{\i}a through the project FQM219. R.C-R. acknowledges support, at different stages of the elaboration of this work, from the Math Institute of the University of Granada (IEMath-GR), the research project MINECO-FEDER MTM2013-47828-C2-1-P, and the Claude Leon Foundation.


\appendix

\section{Useful identities \label{sec:app}}

In this appendix me make a detailed discussion of the necessary mathematical identities that are necessary in order to deduce the form of the field equations, thus filling the gaps in the discussion presented in Sec. \ref{sec:feq}.

The starting point in dealing with the variational principle leading to the gravitational field equations is to consider the variation of the Ricci tensor defined in Eq. \eqref{eq:rictendef}, namely
\begin{equation}
\delta\bar{R}_{(ab)}=\partial_c\delta\bar{\Gamma}^c_{ab}-\partial_{(b}\delta\bar{\Gamma}^c_{a)c}+\delta\bar{\Gamma}^c_{ce}\bar{\Gamma}^e_{ab}+\bar{\Gamma}^c_{ce}\delta\bar{\Gamma}^e_{ab}-
\delta\bar{\Gamma}^c_{be}\bar{\Gamma}^e_{ac}-\bar{\Gamma}^c_{be}\delta\bar{\Gamma}^e_{ac}.
\end{equation}
This equation can be conveniently written as
\begin{equation}
\delta \bar{R}_{(ab)}=\bar{\nabla}_c\delta\bar{\Gamma}^c_{ab}-\bar{\nabla}_{(a}\delta\bar{\Gamma}^c_{b)c}=-\delta^{de}_{c(a}\bar{\nabla}_e\delta\bar{\Gamma}^c_{b)d},
\end{equation}
using the definition of the generalized Kronecker delta
\begin{equation}
\delta^{de}_{ac}=2\delta^{[d}_a\delta^{e]}_c.
\end{equation}
The variation of the gravitational Lagrangian density defined in Eq. \eqref{eq:glagrangian} is then given by
\begin{equation}
2\kappa\,\delta\mathscr{L}_g=\sqrt{-\omega}\,(g/\omega)^{1/4}\left[\bar{R}_{(ab)}-\frac{1}{4}\bar{R}\,g_{ab}\right]\delta g^{ab}+\sqrt{-\omega}\,(g/\omega)^{1/4}g^{ab}\delta^{de}_{ac}\bar{\nabla}_e\delta\bar{\Gamma}^c_{db}.\label{eq:int1app}
\end{equation}
While it is straightforward to obtain the corresponding piece of the gravitational field equations from the first term in the previous equation, doing the same with the last term needs more elaboration. Note that in the Einstein-Hilbert case the corresponding term, coming from the variation of the Ricci tensor, is merely a boundary term thus making things simpler.

The corresponding piece can be written as
\begin{align}
&\sqrt{-\omega}\,(g/\omega)^{1/4}g^{ab}\delta^{de}_{ac}\bar{\nabla}_e\delta\bar{\Gamma}^c_{db}\nonumber\\
&=\sqrt{-g}\,\bar{\nabla}_e\left[(\omega/g)^{1/4}g^{ab}\delta^{de}_{ac}\delta\bar{\Gamma}^c_{db}\right]-\sqrt{-g}\,\delta^{de}_{ac}\delta\bar{\Gamma}^c_{db}\bar{\nabla}_e\left[(\omega/g)^{1/4}g^{ab}\right].\label{eq:int2}
\end{align}
Using the relation
\begin{equation}
\bar{\nabla}_aX^a=\nabla_aX^a-2\mathcal{Q}A_aX^a,
\end{equation}
where $X^a$ is an arbitrary vector field, the first term of Eq. \eqref{eq:int2} can be written, up to boundary terms, as
\begin{align}
&\sqrt{-\omega}\,(g/\omega)^{1/4}g^{ab}\delta^{de}_{ac}\bar{\nabla}_e\delta\bar{\Gamma}^c_{db}\nonumber\\
&=-2\mathcal{Q}\sqrt{-g}\,A_e(\omega/g)^{1/4}g^{ab}\delta^{de}_{ac}\delta\bar{\Gamma}^c_{db}-\sqrt{-g}\,\delta^{de}_{ac}\delta\bar{\Gamma}^c_{db}\bar{\nabla}_e\left[(\omega/g)^{1/4}g^{ab}\right]\nonumber\\
&=-\sqrt{-g}\,\delta^{de}_{ac}\delta\bar{\Gamma}^c_{db}\left(\bar{\nabla}_e\left[(\omega/g)^{1/4}g^{ab}\right]+2\mathcal{Q}A_e\,(\omega/g)^{1/4}g^{ab}\right).\label{eq:int3}
\end{align}
Let us take a detour to show that this expression is invariant under Weyl transformations. The transformation of Christoffel symbols under an infinitesimal transformation $g_{ab}\longrightarrow g_{ab}+\delta g_{ab}$ is given by
\begin{equation}
\delta\bar{\Gamma}^c_{ab}=\frac{1}{2}g^{cd}\left(\nabla_a\delta g_{db}+\nabla_b\delta g_{ad}-\nabla_d\delta g_{ab}\right)+\frac{\mathcal{Q}}{2}A_d\,\delta(g_{ab}g^{cd}),\label{eq:int4app}
\end{equation}
where $\nabla_a$ is the covariant derivative compatible with $g_{ab}$. The first term in the latter equation is the infinitesimal difference $\delta\Gamma^c_{ab}$ between the Levi-Civita connections associated with $g_{ab}+\delta g_{ab}$ and $g_{ab}$. Being the difference between two Weyl-invariant objects, namely two Weyl connections, Eq. \eqref{eq:int4app} must be invariant under Weyl transformations \eqref{eq:wtrans}. This can be checked explicitly, writing
\begin{equation}
\delta\bar{\Gamma}^c_{ab}=\frac{1}{2}g^{cd}\left(\partial_a\delta g_{db}+\partial_b\delta g_{ad}-\partial_d\delta g_{ab}\right)-\Gamma^e_{ab}g^{cd}\delta g_{ed}+\frac{\mathcal{Q}}{2}A_d\,\delta(g_{ab}g^{cd}).\label{eq:intp}
\end{equation}
Under a Weyl transformation \eqref{eq:wtrans},
\begin{align}
\delta\bar{\Gamma}^c_{ab}\longrightarrow\ &\delta\bar{\Gamma}^c_{ab}+g^{cd}\left(\delta g_{db}\partial_a\ln\Omega+\delta g_{ad}\partial_b\ln\Omega-\delta g_{ab}\partial_d\ln\Omega\right)\nonumber\\
&-g^{cd}\left(\delta^e_a\partial_b\ln\Omega+\delta^e_b\partial_a\ln\Omega-g_{ab}\partial^e\ln\Omega\right)\delta g_{ed}+\partial_d\ln\Omega\,\delta(g_{ab}g^{cd}).
\end{align}
Taking into account that $g^{cd}\delta g_{de}=-g_{de}\delta g^{cd}$, it results indeed
\begin{equation}
\delta\bar{\Gamma}^c_{ab}\longrightarrow\delta\bar{\Gamma}^c_{ab}.
\end{equation}
Let us write Eq. \eqref{eq:intp} in a compact notation as
\begin{equation}
\delta\bar{\Gamma}^c_{db}=-\frac{1}{2}D^{fpq}_{\ \ \ dbl}g^{cl}g_{pr}g_{qs}\nabla_f\delta g^{rs}+\frac{\mathcal{Q}}{2}(A_rg_{db}\delta^c_s-A^cg_{dr}g_{bs})\delta g^{rs},
\end{equation}
with
\begin{equation}
D^{fpq}_{\ \ \ dbl}=\delta^f_d\delta_l^{(p}\delta_b^{q)}+\delta^f_b\delta_l^{(p}\delta_d^{q)}-\delta^f_l\delta^{(p}_d\delta^{q)}_b.
\end{equation}
Therefore the last term in Eq. \eqref{eq:int1app} is given by
\begin{align}
&\sqrt{-\omega}\,(g/\omega)^{1/4}g^{ab}\delta^{de}_{ac}\bar{\nabla}_e\delta\bar{\Gamma}^c_{db}\nonumber\\
&=\frac{1}{2}\sqrt{-g}\,\delta^{de}_{ac}D^{fpq}_{\ \ \ dbl}g^{cl}g_{pr}g_{qs}\left(\bar{\nabla}_e\left[(\omega/g)^{1/4}g^{ab}\right]+2\mathcal{Q}A_e\,(\omega/g)^{1/4}g^{ab}\right)\nabla_f\delta g^{rs}\nonumber\\
&-\frac{\mathcal{Q}}{2}\sqrt{-g}\,\delta^{de}_{ac}(A_rg_{db}\delta^c_s-A^cg_{dr}g_{bs})\left(\bar{\nabla}_e\left[(\omega/g)^{1/4}g^{ab}\right]+2\mathcal{Q}A_e\,(\omega/g)^{1/4}g^{ab}\right)\delta g^{rs}.\label{eq:int3}
\end{align}
Note that
\begin{equation}
\delta^{de}_{ac}D^{fpq}_{\ \ \ dbl}=E^{efpq}_{\ \ \ \ acbl}=2\delta^f_{[a}\delta^e_{c]}\delta^{(p}_l\delta^{q)}_b+2\delta^f_b\delta^{(p}_{[a}\delta^e_{c]}\delta^{q)}_l+2\delta^f_l\delta^e_{[a}\delta^{(p}_{c]}\delta^{q)}_b.
\end{equation}
Eq. \eqref{eq:int3} permits to obtain the form of the corresponding piece $\Delta_{ab}$ in the field equations,
\begin{equation}
\bar{R}_{(ab)}-\frac{1}{4}\bar{R}\,g_{ab}+\Delta_{ab}=\kappa\,(g/\omega)^{1/4}\,T^A_{ab},
\end{equation}
as
\begin{align}
\Delta_{ab}=&-\frac{1}{2}(g/\omega)^{1/4}\,E^{efpq}_{\ \ \ \ rcsl}g^{cl}g_{pa}g_{qb}\nabla_f\left(\bar{\nabla}_e\left[(\omega/g)^{1/4}g^{rs}\right]+2\mathcal{Q}A_e\,(\omega/g)^{1/4}g^{rs}\right)\nonumber\\
&-\frac{\mathcal{Q}}{2}(g/\omega)^{1/4}\,\delta^{de}_{rc}\left(A_{(a}\delta^c_{b)}g_{ds}-A^cg_{d(a}g_{b)s}\right)\left(\bar{\nabla}_e\left[(\omega/g)^{1/4}g^{rs}\right]+2\mathcal{Q}A_e\,(\omega/g)^{1/4}g^{rs}\right).\label{eq:deltadef}
\end{align}
Using the contravariant version of Eq. \eqref{eq:nocomp}, this tensor can also be written as
\begin{align}
\Delta_{ab}=&-\frac{1}{2}(g/\omega)^{1/4}\,E^{efpq}_{\ \ \ \ rcsl}g^{cl}g_{pa}g_{qb}\nabla_f\left(\nabla_e\left[(\omega/g)^{1/4}g^{rs}\right]+\mathcal{Q}A_e\,(\omega/g)^{1/4}g^{rs}\right)\nonumber\\
&-\frac{\mathcal{Q}}{2}(g/\omega)^{1/4}\,\delta^{de}_{rc}\left(A_{(a}\delta^c_{b)}g_{ds}-A^cg_{d(a}g_{b)s}\right)\left(\nabla_e\left[(\omega/g)^{1/4}g^{rs}\right]+\mathcal{Q}A_e\,(\omega/g)^{1/4}g^{rs}\right).
\end{align}
Using the following identities,
\begin{equation}
E^{efpq}_{\ \ \ \ rcsl}g^{cl}g^{rs}g_{pa}g_{qb}=2\left( \delta^e_{(a}\delta^f_{b)}-g_{ab}g^{ef}\right),
\end{equation}
\begin{equation}
\delta^{de}_{rc}g^{rs}g_{ds}=3\delta^e_c,
\end{equation}
and
\begin{equation}
\delta^{de}_{rc}g_{d(a}g_{b)s}g^{rs}=g_{ab}\delta^e_c-g_{c(a}\delta^e_{b)},
\end{equation}
Using these, the form of $\Delta_{ab}$ can be further simplified down to
\begin{align}
\Delta_{ab}=&-(g/\omega)^{1/4}\left(\delta^e_{(a}\delta^f_{b)}-g^{df}g_{ab}\right)\nabla_f\left[\partial_e(\omega/g)^{1/4}+\mathcal{Q}A_e\,(\omega/g)^{1/4}\right]\nonumber\\
&-2\mathcal{Q}\left(A_{(a}\delta^e_{b)}-\frac{1}{4}A^eg_{ab}\right)\left[\partial_e\ln(\omega/g)^{1/4}+\mathcal{Q}A_e\right].
\end{align}
It is useful to split this equation in its traceless and pure trace parts. This decomposition is given by
\begin{align}
\Delta_{ab}=&-(g/\omega)^{1/4}\left(\delta^e_{(a}\delta^f_{b)}-\frac{1}{4}g^{df}g_{ab}\right)\nabla_f\left[\partial_e(\omega/g)^{1/4}+\mathcal{Q}A_e\,(\omega/g)^{1/4}\right]\nonumber\\
&-2\mathcal{Q}\left(A_{(a}\delta^e_{b)}-\frac{1}{4}A^eg_{ab}\right)\left[\partial_e\ln(\omega/g)^{1/4}+\mathcal{Q}A_e\right]\nonumber\\
&+\frac{3}{4}g_{ab}(g/\omega)^{1/4}g^{cd}\nabla_c\left[\partial_d(\omega/g)^{1/4}+\mathcal{Q}A_d\,(\omega/g)^{1/4}\right].
\end{align}
Using the definition of the covariant derivative $\bar{\nabla}$, the previous equation can be equivalently written as
\begin{align}
\Delta_{ab}=&-(g/\omega)^{1/4}\left(\delta^e_{(a}\delta^f_{b)}-\frac{1}{4}g^{df}g_{ab}\right)\bar{\nabla}_f\left[\partial_e(\omega/g)^{1/4}+\mathcal{Q}A_e\,(\omega/g)^{1/4}\right]\nonumber\\
&-\mathcal{Q}\left(A_{(a}\delta^e_{b)}-\frac{1}{4}A^eg_{ab}\right)\left[\partial_e\ln(\omega/g)^{1/4}+\mathcal{Q}A_e\right]\nonumber\\
&+\frac{3}{4}g_{ab}(g/\omega)^{1/4}g^{cd}\nabla_c\left[\partial_d(\omega/g)^{1/4}+\mathcal{Q}A_d\,(\omega/g)^{1/4}\right].
\end{align}
This expression makes transparent that $\Delta_{ab}$ is invariant under Weyl transformations \eqref{eq:wtrans}.

Now let us deal with the contribution of the variation of the Ricci tensor in the gravitational Lagrangian density to the electromagnetic field equations. The corresponding terms are given by Eq. \eqref{eq:int3}, namely
\begin{align}
&\sqrt{-\omega}\,(g/\omega)^{1/4}g^{ab}\delta^{de}_{ac}\bar{\nabla}_e\delta\bar{\Gamma}^c_{db}\nonumber\\
&=-\sqrt{-g}\,\delta^{de}_{ac}\delta\bar{\Gamma}^c_{db}\left\{\nabla_e\left[(\omega/g)^{1/4}g^{ab}\right]+\mathcal{Q}A_e\,(\omega/g)^{1/4}g^{ab}\right\},
\end{align}
where now the variations $\delta\bar{\Gamma}^c_{eb}$ with respect to $A_a$ are to be considered. 
\begin{equation}
\delta\bar{\Gamma}^c_{db}=-\frac{\mathcal{Q}}{2}O^{cr}_{\,\,\, db}\delta A_r,
\end{equation}
where
\begin{equation}
O^{cr}_{\,\,\, db}=\delta^c_d\delta^r_b+\delta^c_b\delta^r_d-g_{db}g^{cr},
\end{equation}
so that
\begin{align}
&\sqrt{-\omega}\,(g/\omega)^{1/4}g^{ab}\delta^{de}_{ac}\bar{\nabla}_e\delta\bar{\Gamma}^c_{db}\nonumber\\
&=\frac{\mathcal{Q}}{2}\sqrt{-g}\,\delta^{de}_{ac}O^{cr}_{\,\,\, db}\left\{\nabla_e\left[(\omega/g)^{1/4}g^{ab}\right]+\mathcal{Q}A_e\,(\omega/g)^{1/4}g^{ab}\right\}\delta A_r.
\end{align}
The corresponding piece takes the form of an additional, Weyl invariant current:
\begin{equation}
\nabla_bF^{ab}+3\mathcal{Q}(\omega/g)^{1/4}g^{ab}\left[\partial_b\ln(\omega/g)^{1/4}+\mathcal{Q}A_b\right]=0.
\end{equation}
%

\section{Linear Weyl relativity over a flat background \label{sec:flat}}

As a complementary discussion, let us construct here a linear Weyl-invariant theory for fields $h_{ab}$ and $A_a$ over a Minkowski background. One can easily realize that any combination
\begin{eqnarray}
\partial_c h_{ab} -{\cal Q} A_c \eta_{ab}
\label{eq:linearcov}
\end{eqnarray}
exhibits Weyl invariance, now meaning invariance under $h_{ab} \to h_{ab}+\varphi \eta_{ab}$ and $A_a \to A_a +(1/{\cal Q})\partial_a \varphi$. Now let us consider the invariance of the action under a gauge transformation of the form $h_{ab} \to h_{ab}+\partial_{(a}\xi_{b)}$, hereafter LDiff symmetry. We have choosen this name because this symmetry is equal to the linear version of the diffeomorphism symmetry group Diff in standard general relativity. Recall however that at the linear level this symmetry has nothing to do with changes of coordinates. Note also that at the linear level we are not requiring the modification of $A_a$ under LDiff.

Following~\cite{Alvarezetal2006} we know that there is a family of systems that are invariant under transverse transformations in LDiff (TLDiff), that is, those LDiff transformations satisfying $\partial_a \xi^a=0$. If we define the following possible terms that a linear theory might have
\begin{align}
&\mathscr{L}_1 =-{1 \over 4} \partial_a h^{bc} \partial^a h_{bc}, &\mathscr{L}_2 =-{1 \over 2} \partial_a h^{ab} \partial_c h^{c}_{~b},
\\
&\mathscr{L}_3 ={1 \over 2} \partial_a h^{ab} \partial_b h, &\mathscr{L}_4 =-{1 \over 4} \partial_a h \partial^a h,
\label{eq:Lagrangians}
\end{align}
it is possible to show that any combination 
\begin{eqnarray}
\mathscr{L}=\mathscr{L}_1 - \mathscr{L}_2 +c_3 \mathscr{L}_3 +c_4 \mathscr{L}_4,
\label{eq:Lagrangiancombination}
\end{eqnarray}
with $c_3,c_4$ arbitrary constants is TLDiff invariant. To construct a Weyl invariant theory we just have to substitute $\partial_c h_{ab} \to \partial_c h_{ab}-{\cal Q}A_c\eta_{ab}$. However, in doing so one has to ensure that the new interaction terms between $h_{ab}$ and $A_a$ do not spoil the TLDiff symmetry of the free kinetic terms. It is not difficult to see that this condition fixes $c_3=-1/2$, but leaving $c_4$ still arbitrary. 

The linear Weyl tranverse gravity Lagrangian in~\cite{Alvarezetal2006} fixes also the value of $c_4=-3/8$. Thus, the linear Weyl relativity theory we are constructing here, when $A_a=0$ can be written as   
\begin{eqnarray}
\mathscr{L}=\mathscr{L}_{\rm WTgrav} + (c_4+3/8) \mathscr{L}_4.
\label{eq:Lagrangian-grav}
\end{eqnarray}
Whenever $c_4\neq -3/8$ this gravitational theory has one degree of freedom more that WTDiff gravity and Fierz-Pauli theory: the trace of $h_{ab}$, a scalar field, is also a dynamical piece of the theory. Thus, the theory entails spin-two gravitons plus a scalar field. It is also interesting to realize that already at the linear level it is not possible to have the entire LDiff symmetry and Weyl invariance: starting from a Fierz-Pauli Lagrangian the interaction terms necessary for adding Weyl invariance break the initial LDiff invariance.

The entire linear Weyl relativity Lagrangian we have built reads
\begin{eqnarray}
\mathscr{L}=\mathscr{L}_{\rm WTgrav}(\partial_ch_{ab}-A_c\eta_{ab}) + (c_4+3/8) \mathscr{L}_4(\partial_ch_{ab}-A_c\eta_{ab}) -{1 \over 4} F_{ab}F^{ab}.
\label{eq:Lagrangian-Weyl-rel}
\end{eqnarray}
In fact, it is not difficult to check that this theory can be written equivalently in the simpler fashion 
\begin{eqnarray}
\mathscr{L}=\mathscr{L}_{\rm WTgrav}(\partial_c h_{ab}) + (c_4+3/8) \mathscr{L}_4(\partial_c h_{ab}-A_c\eta_{ab}) -{1 \over 4} F_{ab}F^{ab}.
\label{eq:Lagrangian-Weyl-rel}
\end{eqnarray}
The $\mathscr{L}_{\rm WTgrav}(\partial_ch_{ab})$ piece is Weyl invariant by itself without the need to add to it interaction terms. Apart from this term, the rest of the Lagrangian 
\begin{eqnarray}
-{1 \over 4}(c_4+3/8) (\partial_a h-4{\cal Q} A_a) (\partial^a h-4{\cal Q} A^a) -{1 \over 4} F_{ab}F^{ab},
\label{eq:Lagrangian-Stueckelberg}
\end{eqnarray}
can be seen to be equivalent to the Stueckelberg Lagrangian \cite{Ruegg2003}. By defining a new $\bar{A}_a$ as 
\begin{eqnarray}
\bar{A}_a= A_a - {1 \over 4 {\cal Q}} \partial_a h,
\label{eq:Lagrangian-Stueckelberg-trick}
\end{eqnarray}
this remaining piece of the theory can be seen to correspond to a Proca field. Therefore the new scalar degree of freedom in the gravitational theory can be seen alternatively as a longitudinal degree of freedom for the electromagnetic field. The geometric theory we have presented in the bulk of the paper corresponds in the linear limit to this theory precisely.

In the linear analysis just above one can easily distinguish between invariance under changes of coordinates (Diff) and gauge transformations (LDiff). Being the background metric non-dynamical, Diff invariance is not fully implemented though. By itself the invariance under coordinate changes does not play any role in eliminating degrees of freedom as gauge invariance does. However, at the non-linear level these two notions become mixed up. As is well known in standard general relativity, the deformation of LDiff leads to proper Diff in a curved and dynamical manifold \cite{Deser1969,Barcelo2014,Ortin2015}.

\bibliography{weyl}
\end{document}